\documentclass[prd, onecolumn, notitlepage, nofootinbib,showkeys, amsmath, amssymb, superscriptaddress]{revtex4-1}

\usepackage{graphicx}
\usepackage{dcolumn}
\usepackage{bm}
\usepackage{xcolor}
\usepackage{hyperref}
\hypersetup{
    colorlinks = true, 
    linkcolor = black,
    citecolor = black,
    urlcolor = teal,
}
\usepackage{lipsum}
\usepackage{float}
\usepackage{mathrsfs}

\usepackage[T1]{fontenc}
\usepackage[latin9]{inputenc}

\usepackage{tabularx}
\newcolumntype{Y}{>{\centering\arraybackslash}X}
\usepackage{multirow}

\usepackage{ulem}

\graphicspath{{Images/}}

\begin{document}

\title{Dynamical analysis of null geodesics in brane-world spacetimes}

\author{W. S. Kl\"{e}n}
\email{wayner.klen@usp.br}

\author{C. Molina}
\email{cmolina@usp.br}

\affiliation{Escola de Artes, Ci\^{e}ncias e Humanidades, Universidade de S\~{a}o Paulo. \\
Av. Arlindo B\'{e}ttio 1000, CEP 03828-000, S\~{a}o Paulo-SP, Brazil
}

\begin{abstract}
In this work, we present an extensive dynamical analysis of the null geodesics in the spacetimes proposed by Casadio, Fabbri, and Mazzacurati (CFM), which model black holes and wormholes in a Randall-Sundrum brane. Geodesic stability is evaluated by Lyapunov and Jacobi criteria, with coinciding results. Semistable photon spheres are found in the geometries of interest. Bifurcations associated with the dynamical system are also investigated, characterizing a degenerated Bogdanov-Takens bifurcation. Our results suggest deep connections between the geometric characteristics of the CFM spacetimes and the dynamics of null geodesics in these backgrounds.
\end{abstract}


\maketitle

\section{\label{sec:Intro}Introduction}

Randall-Sundrum brane models \cite{randall1999large,randall1999alternative} are concrete implementations of extra-dimensional scenarios predicted by string theories. Although exact bulk solutions describing compact objects in Randall-Sundrum branes are not easily obtained, new insights were possible with four-dimensional effective theories in the brane \cite{shiromizu2000einstein}. Within this formalism, brane black hole and wormhole geometries were proposed 
\cite{Dadhich:2000am,casadio2002new,Bronnikov:2003gx,Lobo:2007qi,molina2010black,molina2012wormholes,Neves:2012it,Neves:2014aba,Molina:2016tkr}.
The investigation of the geodesic dynamics around compact objects in brane scenarios could open a new window for the physics beyond standard general relativity. For instance, the recent direct observations of gravitational waves \cite{Abbott:2016blz} in coalescing black holes are generating data which can be compared with theoretical predictions. More recently, the first image of a black hole 
\cite{event2019firstI} has also yielded promising results. In both the black hole merging and the event horizon imaging, geodesic dynamics is relevant.

Considering the set of ordinary differential equations that describe the geodesic dynamics, the framework of qualitative dynamic theory can get relevant information about the system.
Usually, the main goal in this analysis is to understand the asymptotic behavior of the solutions, and for that, the canonical procedure is the evaluation of quantities which characterizes the stability, such as Lyapunov exponents \cite{bombelli1992chaos,cornish2003lyapunov,cardoso2009geodesic}. 
However, there are other concepts of stability which can be used to understand the dynamics, such as the Jacobi stability criterion \cite{sabau2005some,boehmer2012jacobi}. From a general point of view, some systems display concordance between both concepts of stability and these are of special interest once this concordance can be interpreted as an indicator of a robust stability \cite{sabau2005some,abolghasem2013stability}.
The nonlinear nature of Einstein's field equations gives rise to rich dynamical characteristics to its solutions, such as bifurcations, special kinds of orbits like homoclinic or heteroclinic and even the emergence of chaos. The tracking of these structures shows the possibility of new behaviors to be expected through observations.

The so-called photon surfaces (or light surfaces) are of considerable interest in the present work. These structures are regions where light is confined in closed orbits \cite{Claudel:2000yi,Foertsch:2003ze}. The connection between photon-surface parameters and quantities associated with the perturbative dynamics around black holes has been recently explored, for example, in \cite{cardoso2009geodesic,PhysRevD.81.104039,PhysRevD.90.044069}. A semiclassical treatment of the photon surface was presented in \cite{Baldiotti:2014pca}, in the context of spherically symmetric and static geometries (in this case, the photon surface is a photon sphere). From the observational point of view, the photon-surface phenomenology is also an issue, and their astrophysical signatures were discussed in \cite{Moscibrodzka:2009gw,Johannsen:2010ru,PhysRevD.90.044069}. 
It should be remarked that, although black-hole scenarios are better explored within the current observations, wormholes are not excluded \cite{Damour:2007ap, Cardoso:2016rao}.

In the present work, a dynamical analysis is performed considering black holes and wormholes in the family of geometries proposed by Casadio, Fabbri and Mazzacurati (CFM) \cite{casadio2002new}. The CFM geometries can be interpreted as four-dimensional brane solutions in a Randall-Sundrum scenario, and could be understood as a deformation of the usual Schwarzschild spacetime 
\cite{Bronnikov:2003gx,Molina:2011mc,Molina:2013mwa,Molina:2016tkr}.
Null geodesics around CFM black holes and wormholes are studied and their dynamics explored.
Characteristics of the dynamical system of interest, such as bifurcations and the existence of heteroclinic orbits, are investigated.

The structure of this paper is described in the following. 
In section~\ref{sec:brane-sol}, the brane spacetimes considered are introduced and discussed. 
In section~\ref{sec:photon-spheres}, the effective dynamical system is constructed. Relevant features of this system, such as the existence of semistable photon spheres, are highlighted.
A stability analysis is performed in section~\ref{sec:sta-ana} with a variety of approaches. 
In section~\ref{sec:bifurcations} bifurcations in the dynamics are treated in detail. 
The existence of heteroclinic orbits is considered in section~\ref{sub:heteroclinic} and 
final remarks are presented in section~\ref{sec:conclusions}.
In the present work we use metric signature $\textrm{diag}(+---)$ and geometric units with $G_{4D}=c=1$, where $G_{4D}$ is the effective four-dimensional gravitational constant.

\section{\label{sec:brane-sol} CFM spacetimes and geodesics}

The main characteristics of brane-world models are that the matter fields are confined to a four-dimensional hypersurface, the brane, while gravity propagates in a larger spacetime, the bulk \cite{Maartens:2010ar}. Brane worlds where the extra dimension is noncompact can be implemented, and we focus on the Randall-Sundrum setup, where our universe is a domain wall embedded in a five-dimensional anti--de Sitter spacetime \cite{randall1999large,randall1999alternative}.

While complete brane-world cosmological solutions (including both the bulk and the brane) are relatively common, the derivation of exact bulk solutions modeling compact objects in Randall-Sundrum scenarios is more difficult. One alternative is to construct four-dimensional geometries in the brane and invoke Campbell-Magaard theorems \cite{Seahra:2003eb}, which guarantees their extensions through the bulk (at least locally). Four-dimensional gravitational field equations obtained by Shiromizu, Maeda and Sasaki \cite{shiromizu2000einstein} are assumed to describe gravity in the brane. The vacuum version of their effective equations, considering asymptotically flat branes, are given by
\begin{equation}
R_{\mu\nu} - \frac{1}{2} R g_{\mu\nu} = - \mathcal{E}_{\mu\nu} \,\, ,
\label{vaccum-Randall-Sundrum}
\end{equation}
where $\mathcal{E}_{\mu \nu}$ is the projection of the five-dimensional Weyl tensor on the brane.
In brane-world models, $\mathcal{E}_{\mu\nu}$ is traceless \cite{Maartens:2010ar}, and hence
\begin{equation}
R = 0 \,\, .
\label{Ricci}
\end{equation}

Considering a spherically symmetric and static four-dimensional brane, the metric has the form given by
\begin{equation}
    \textup{d}s^{2} = A(r)\textup{d}t^{2} - B(r)^{-1}\textup{d}r^{2} - r^{2}(d\theta^{2} + \sin^{2} \theta d\phi^{2}) \,\, .
    \label{ansatz}
\end{equation}
In this case, relation~(\ref{Ricci}) can be rewritten as a constraint between the metric functions $A(r)$ and $B(r)$,
\begin{equation}
\frac{A''}{A} - \frac{(A')^{2}}{2A^{2}} + \frac{A'B'}{2AB} + \frac{2}{r} \left[\frac{A'}{A} + \frac{B'}{B}\right]
= \frac{2(1-B)}{r^{2} B} \,\, ,
\label{constraint}
\end{equation}
with prime ($'$) denoting differentiation with respect to $r$. 

We are interested in solutions ``close'' to the Schwarzschild-like geometries, in which $A(r)=B(r)$. A family of spacetimes which are one-parameter continuous deformations of the Schwarzschild spacetime was obtained by Casadio, Fabbri, and Mazzacurati \cite{casadio2002new,Molina:2016tkr,Molina:2013mwa}. The CFM metric can be written as
\begin{equation}
A(r) = 1-\frac{2M}{r} \,\, , \,\, 
B(r) = A(r)\left[1 + \frac{(C-1)M}{2r-3M}\right]  \,\, .
\label{NCFM}
\end{equation}
The constant $M$ is the Misner-Sharp mass of the compact object, while $C$ can be interpreted as a deformation parameter. It is assumed that $M$ is positive and $C \in \mathbb{R}$.
The real and positive quantity $r_{+} \equiv 2M$ is a root of both $A(r)$ and $B(r)$. If $C<0$, the function $B(r)$ has an additional positive root $r_{\textrm{thr}}$, with $r_{\textrm{thr}} > r_{+}$.
It should be remarked that the coordinate system $(t,r,\theta,\phi)$ is valid for the radial coordinate $r$ taking values in a subset of $\mathbb{R}$ such that $A(r)>0$ and $B(r)>0$.

Considering the asymptotic limit $r \rightarrow \infty$ of the metric functions, the CFM metric functions behave as
\begin{equation}
A(r) \sim B(r) \sim 1 + \mathcal{O} \left( \frac{1}{r} \right) \,\, .
\end{equation}
Hence the CFM spacetimes are asymptotically flat for any value of the deformation parameter $C$. But further global properties depend strongly on $C$.

For instance, if $C \ge 0$, the analytic extension beyond $r=r_{+}$ can be performed with the usual methods, for example using the ingoing and outgoing Eddington charts. In the maximal extension, if $C>0$ the surface $r=r_{+}$ is an event horizon with a nonvanishing surface gravity. When $C=0$, the surface gravity of the horizon is zero.
The CFM spacetime with $C \ge 1$ describes a black hole with a singularity hidden by the event horizon \cite{casadio2002new,Molina:2016tkr}. In the specific case $C=1$ the usual Schwarzschild solution is recovered.
If $0 < C < 1$, the CFM solution still describes a black hole surrounded by an event horizon at $r=r_{+}$. But its interior is regular, with no singularity \cite{casadio2002new,Molina:2016tkr}.
If $C=0$, the derived solution models an extreme black hole \cite{casadio2002new,Molina:2016tkr}.
For non-negative values of the deformation parameter $C$, the coordinate system $(t,r,\theta,\phi)$ is valid when $r>r_{+}$.

For negative values of $C$, the CFM metric no longer describes a black hole, but instead a symmetric and transversable wormhole. In this case, the function $B(r)$ has two simple positive zeros, $r_{+}$ and $r_{\textrm{thr}}$. The zero $r_{+}$ does not play any role in the causal structure since $r_{+} < r_{\textrm{thr}}$. In its maximal extension, the CFM geometry has a wormhole structure with a throat at $r \rightarrow r_{\textrm{thr}}$ \cite{casadio2002new,Molina:2016tkr}.
With $C<0$, the coordinate system $(t,r,\theta,\phi)$ is valid when $r>r_{\textrm{thr}}$. The maximal extension can be made changing the ``areal'' coordinate $r$ to the quasiglobal coordinate $u$ \cite{Bronnikov:2008by}, where
\begin{equation}
\frac{\textup{d}u}{\textup{d}r} = \sqrt{\frac{A(r)}{B(r)}} \,\, .
\label{def-u}
\end{equation}
The chart $(t,u,\theta,\phi)$ with $-\infty < u < \infty$ maps the whole wormhole spacetime. With a convenient choice of the integration constant in Eq.~\eqref{def-u}, the throat is located at $u=0$.

With the main characteristics of the CFM spacetimes reviewed, the next step is to consider the propagation of massless particles in these backgrounds. In order to derive the geodesic equations of motion, let us consider the following action \cite{chandrasekhar1985mathematical}:
\begin{equation}
    I = \int \mathscr{L}\ \textup{d}\lambda \,\, , 
\quad \textrm{with} \quad 2\mathscr{L} = g_{\mu \nu}\frac{\textup{d}x^{\mu}}{\textup{d} \lambda}\frac{\textup{d}x^{\nu}}{\textup{d} \lambda} \,\, .
   \label{action}
\end{equation}
Extremizing the action~\eqref{action} we get the Euler-Lagrange equations for the problem associated with the functional $I$, and the geodesic equation is obtained \cite{chandrasekhar1985mathematical}.
By evaluating the associated Hamiltonian $\mathscr{H}$ we find that $\mathscr{L} = \mathscr{H} = \textrm{constant}$, and
\begin{equation}
2 \mathscr{H}
= \frac{p_{t}^{2}}{A(r)} - B(r) \, p_{r}^{2} - \frac{p_{\theta}^{2}}{r^{2}} - \frac{p_{\phi}^{2}}{r^{2} \sin^{2}\theta}
\,\, ,
\label{hamiltonian}
\end{equation}
where the conjugated momenta read 
\begin{equation}
p_{t} = A(r) \, \frac{\textup{d}t}{\textup{d}\lambda} \,\, , \,
p_{\theta} = - r^2 \, \frac{\textup{d}\theta}{\textup{d}\lambda} \,\, , \,
p_{\phi} = - r^2 \sin^{2} \theta \, \frac{\textup{d}\phi}{\textup{d}\lambda} \,\, , \,
p_{r} = - B(r)^{-1} \, \frac{\textup{d}r}{\textup{d}\lambda} \,\, .
\label{conju_moments}
\end{equation}

Hamilton's equations of motion $\textup{d} p_{\mu} / \textup{d} \lambda = \partial \mathscr{H}/ \partial x^{\mu}$ applied to the components $t$ and $\phi$ imply that
\begin{equation}
p_{t} = E = \textrm{constant} \,\, , \, p_{\phi} = L = \textrm{constant} \,\, .
\label{conserved}
\end{equation}
The integration constants $E$ and $L$ are related to the energy and angular momentum of a given geodesic. The existence of these conserved quantities is a consequence of the staticity and spherical symmetry of the geometry.
Also, from Hamilton's equation $\textup{d} p_{\theta} / \textup{d} \lambda = \partial \mathscr{H}/ \partial \theta$, it is observed that the geodesic is restricted to an invariant plane which may be chosen as $\theta = \pi/2$.
Considering null geodesics ($\mathscr{H}=0$) and manipulating the Hamiltonian~\eqref{hamiltonian}, one gets the equation of motion for the radial component, 
\begin{equation}
 \left(\frac{\textup{d}r}{\textup{d}\lambda}\right)^{2} = E^{2}\left[\frac{B(r)}{A(r)}-B(r) \, \frac{D^{2}}{r^{2}}\right] \,\, ,
\label{r_dot}
\end{equation}
where the impact parameter $D$ is defined as
\begin{equation}
D \equiv \frac{L}{E} \,\, .
\end{equation}
Introducing a  new affine parameter $\eta$ defined as
\begin{equation}
\eta \equiv E \lambda \,\, ,
\end{equation}
the explicit mention to $E$ is eliminated from the system, and Eq.~\eqref{r_dot} is written as
\begin{equation}
 \left(\frac{\textup{d}r}{\textup{d}\eta}\right)^{2} = \frac{B(r)}{A(r)}-B(r) \, \frac{D^{2}}{r^{2}} \,\, .
 \label{r_dot_eta}
\end{equation}

The differential equation~\eqref{r_dot_eta}, together with relations~\eqref{conserved} restricted (without loss of generality) to the plane $\theta = \pi/2$,
\begin{align}
\frac{\textup{d}t}{\textup{d}\eta} &= \frac{1}{A(r)} \,\, , 
\label{eq-t-eta}  \\
\frac{\textup{d}\phi}{\textup{d}\eta} &=- \frac{D}{r^{2}} \,\, ,
\label{eq-phi-eta}
\end{align}
form the extended set of differential equations that characterize the null-geodesic dynamics in spherically symmetric and static spacetimes.
In this extended system, Eq.~\eqref{r_dot_eta} is decoupled, and with $r(\eta)$ determined the functions $t(\eta)$ and $\phi(\eta)$ can be calculated in a straightforward way.

Although Eqs.~\eqref{r_dot_eta}-\eqref{eq-phi-eta} completely  define the dynamics of the null geodesics in spherically symmetric and static spacetimes, this is not the most convenient way of treating the problem considering the developments to be done in the present work. In the next sections, a new version of the dynamical system will be introduced and explored.

\section{\label{sec:photon-spheres}Effective dynamics and photon spheres}

Instead of treating the set of differential equations~\eqref{r_dot_eta}-\eqref{eq-phi-eta}, it is more convenient to work with an effective dynamical system, associated only with its radial dependence.
Differentiating relation~\eqref{r_dot_eta} with respect to $\eta$,  the obtained autonomous second-order differential equation is
\begin{equation}
    \frac{\textup{d}^{2}r}{\textup{d}\eta^{2}} = \frac{f'(r)}{2} \,\, ,
    \label{geodesic_equation}
\end{equation}
where the function $f(r)$ is defined as
\begin{equation}
f(r) \equiv \frac{B(r)}{A(r)}-B(r) \, \frac{D^{2}}{r^{2}} \,\, .
\label{def-f}
\end{equation}

From Eq.~\eqref{geodesic_equation}, the effective system associated to the null geodesics is written in the phase portrait $\mathcal{F}$, with $\mathcal{F} \subset \mathbb{R}^{2}$, labeled by the coordinates $(r,w)$, where
\begin{equation}
\frac{\textup{d}r}{\textup{d}\eta} =  w \,\, , \,\,\,\,
\frac{\textup{d}w}{\textup{d}\eta} =  \frac{f'(r)}{2} \,\, .
\label{vec-CFM}
\end{equation}
Also, according to Eq.~\eqref{r_dot_eta}, the first-order differential equations in \eqref{vec-CFM} are subjected to the constraint
\begin{equation}
w^{2} = f(r) \,\, .
\label{geo_const}
\end{equation}

This constrained system can be turned into an unconstrained problem. Relation~\eqref{geo_const} can be used to eliminate $D$ from equations in~\eqref{vec-CFM}, yielding the following unconstrained system:
\begin{eqnarray}
    \frac{\textup{d}r}{\textup{d}\eta} & = &  w \,\, ,     
\label{vec-CFM-1-eff}\\
    \frac{\textup{d}w}{\textup{d}\eta} & = & \tilde{f}(r,w) \,\, ,
\label{vec-CFM-2-eff}
\end{eqnarray}
where
\begin{equation}
\tilde{f}(r,w) \equiv \tilde{g}(r)+\frac{1}{2} w^2\tilde{h}(r) \,\, ,
\label{eq_frw}
\end{equation}
with
\begin{equation}
\tilde{g}(r) \equiv \frac{B(r)}{A(r)}
\left[ \frac{1}{r} -\frac{A'(r)}{2 A(r)} \right] \,\, , \,\,\,
\tilde{h}(r) \equiv \frac{B'(r)}{B(r)}-\frac{2}{r} \,\, .
\label{def-tildeg-tildeh}
\end{equation}
We have obtained a system that is unconstrained and independent
of $E$ and $L$, with the impact parameter $D$ being determined by the initial data.
The set of differential equations~\eqref{vec-CFM-1-eff}--\eqref{vec-CFM-2-eff} defines the dynamical system explored in the present work.

Denoting a point in the phase portrait by $\mathbf{z} \equiv (r,w)$, and defining the function $\mathbf{F}:\mathcal{F} \rightarrow \mathcal{F}$ as
\begin{equation}
\mathbf{F} (\mathbf{z}) = \left(w, \tilde{f}(r,w) \right) \,\, ,
\end{equation}
the dynamical system~\eqref{vec-CFM-1-eff}--\eqref{vec-CFM-2-eff} can be written as
\begin{equation}
\frac{\textup{d} \mathbf{z}}{\textup{d} \eta} = \mathbf{F} (\mathbf{z}) \,\, .
\label{dynamical_system_ger}
\end{equation}
A fixed point $\mathbf{z}^{\star}$ of the system is defined as
\begin{equation}
\mathbf{F}(\mathbf{z}^{\star}) = 0 \,\, .
\label{cond-fixed-point-1}
\end{equation}
In terms of the function $f(r)$ and $\tilde{f}(r,w)$, the fixed-point condition~\eqref{cond-fixed-point-1} is equivalent to
\begin{equation}
w^{\star} = \sqrt{f(r^{\star})} = 0 \,\, ,
\label{cond-fixed-point-2}
\end{equation}
\begin{equation}
\frac{\textup{d}w^{\star}}{\textup{d}\eta} = \tilde{f}(r^{\star},w^{\star}) = 0 \,\, ,
\label{cond-fixed-point-3}
\end{equation}
with $r^{\star}$ and $w^{\star}$ defined by $\mathbf{z}^{\star} \equiv \left( r^{\star}, w^{\star} \right)$. However, since $w^{\star} =0$, the fixed point $\mathbf{z}^{\star}$ can be written as $\mathbf{z}^{\star} = \left( r^{\star}, 0\right)$.

Of considerable importance for the geodesic dynamics is the concept of photon surfaces. Intuitively, they are regions where particles with null rest mass are confined. More precisely, photon  surfaces are defined as hypersurfaces $S$ where any null geodesic initially tangent to $S$ remains tangent to $S$, that is, given a manifold equipped with a metric, a photon surface is an immersed, nowhere spacelike hypersurface $S$ such that for every point $p \in S$ and every null vector $\mathbf{k} \in T_{p}S$ there exists a null geodesic $\gamma: (-\epsilon,\epsilon) \to M$ of $(M,\mathbf{g})$ such that 
$\left. \textup{d} \gamma/ \textup{d}\eta \right|_{0} = \mathbf{k}$ 
with the image of $\gamma$ being a subset of $S$ \cite{Claudel:2000yi}.
In static and spherically symmetric spacetimes, photon surfaces are regions formed by circular null geodesics around the ultracompact object. In this context, photon surfaces are usually denoted photon spheres.
Alternatively, the geodesic radius should be constant at the photon sphere, that is, $r=r^{\star}$, implying that the rate of variation of the radial coordinates is zero.  This condition can be expressed as $f(r^{\star}) =  \tilde{f}(r^{\star},0) = 0$, where $r^{\star}$ can be interpreted as the radius of the photon sphere \cite{cardoso2009geodesic,Khoo:2016xqv}. 
Considering Eqs.~\eqref{cond-fixed-point-1}--\eqref{cond-fixed-point-3}, the photon-sphere condition is written as  $\mathbf{F}(\mathbf{z}^{\star}) = \mathbf{0}$ in the phase portrait $\mathcal{F}$. Hence, photon spheres are fixed points in the dynamic description of null geodesics according to the effective dynamical system~\eqref{vec-CFM-1-eff}--\eqref{vec-CFM-2-eff}.

Let us proceed to the analysis of the fixed points in the null-geodesic dynamics. Inserting the metric functions~\eqref{NCFM} in the expression~\eqref{eq_frw} for $\tilde{f}(r^{\star},0)$,
\begin{equation}
    \tilde{f}(r^{\star},0) = -\frac{(3 M-r^{\star}) [(C-4) M+2 r^{\star}]}{r^{\star} (3 M-2 r^{\star}) (2 M-r^{\star})} \,\, .
    \label{cfm_f(r)}
\end{equation}
Using~\eqref{cfm_f(r)} with $f(r^{\star},0) = 0$, we obtain the set $\{ r^{\star}_{i} \}$ associated to the fixed points $\mathbf{z}^{\star}_{i} = (r^{\star}_{i},0)$ written in terms of the parameters of the geometry:
\begin{eqnarray}
r^{\star}_{1} & = & \frac{M}{2} (4 - C) \,\, , \\
r^{\star}_{2} & = &  3M  \,\, . 
\label{fixed-points}
\end{eqnarray}

But not necessarily both $r^{\star}_{1}$ and $r^{\star}_{2}$ are physically acceptable, taking into account that only in a subset of the spacetime manifold the geometry is static (and hence equipped with a time-like Killing field).
For instance, when the deformation parameter $C$ is non-negative, that is, considering CFM spacetimes describing black holes, $r = r^{\star}_{i}$ is admissible only if $r^{\star}_{i} > r_{+} \equiv 2M$. Since 
\begin{equation}
C \ge 0 \Longrightarrow r^{\star}_{1} \leq 2M \,\, ,
\end{equation}
then $r^{\star}_{2} \equiv 3M$ is the only physically meaningful fixed point for CFM spacetimes with non-negative $C$. 

If the deformation parameter $C$ is negative, meaning that the CFM geometry models a wormhole, the analysis is more involving. The number of fixed points of the system depends on the value of $C$, and bifurcations are possible. These issues will be considered in section~\ref{sec:bifurcations}. One important result to be mentioned here is that
\begin{equation}
r^{\star}_{1} = r_{\textrm{thr}} 
\end{equation}
when $r^{\star}_{1}$ is an acceptable fixed point. It follows that wormhole throats in CFM geometries are photon spheres.
When $C < -2$, it is straightforward to see that $r^{\star}_{2} < r^{\star}_{1}$. But since $r^{\star}_{1} = r_{\textrm{thr}}$, it follows that $r^{\star}_{2} < r_{\textrm{thr}}$ for $C < -2$. Hence $r^{\star}_{2}$ is not associated to a photon sphere in this range of the parameter $C$, because $r \ge r_{\textrm{thr}}$ for any point of the wormhole spacetime.

Considering null geodesics with the impact parameter $D$ set at any of the following critical values,
\begin{eqnarray}
D^{c}_{1} & = &  - \left[ \frac{4 - C}{(-C)^{1/3}}  \right]^{3/2} \, \frac{M}{2}\,\,, \,\, \textrm{associated to $r^{\star}_{1}$} \,\,,  
\label{crit-par-photon-1} \\
D^{c}_{2} & = & 3\sqrt{3}M \,\,, 
\,\, \textrm{associated to $r^{\star}_{2}$} \,\,, 
\label{crit-par-photon-2}
\end{eqnarray}
there will be fixed points satisfying the photon-sphere conditions $f(r^{\star}_{i}) = f'(r^{\star}_{i}) = 0$ \cite{cardoso2009geodesic,Khoo:2016xqv}. 
It should be noticed that the impact parameter $D^{c}_1$ in Eq.~\eqref{crit-par-photon-1} 
is associated to $r^{\star}_{1}$ and hence is only meaningful if $C<0$. For this range of $C$, $D^{c}_1 \in \mathbb{R}$.

The existence and uniqueness of the solutions assure that there are no intersections of the orbits on the phase portrait.
This way, solutions with $D=D^{c}_{i}$ approaching the photon sphere towards or backwards in $\eta$ will spin infinitely around the photon sphere and never reach $r=r^{\star}_{i}$,that is $r \to r^{\star}_{i}$ as $\phi \to \infty$.
Hence the photon sphere is a limit cycle in the orbital plane, that is, considering the extended set of Eqs.~\eqref{r_dot_eta}--\eqref{eq-phi-eta}.

\section{\label{sec:sta-ana}Stability analysis}

In this section, we introduce the main aspects of the geodesic stability analysis from a Lyapunov and Jacobi point of view.
The effective dynamical system~\eqref{vec-CFM-1-eff}--\eqref{vec-CFM-2-eff} can be studied locally by its linear stability. The method is based on the linearization of the dynamics, evaluating the eigenvalues of the Jacobian matrix associated to $\mathbf{F}$. 
If the fixed points of $\mathbf{F}$ are hyperbolic, that is, if they have non-null real parts, the Hartman-Grobman theorem assures the existence of a homeomorphism between the nonlinear phase portrait and its linearized version. For nonhyperbolic fixed points, case-by-case analyses will be conducted.

To introduce the Lyapunov's criterion of stability let us denote the Jacobian matrix of $\mathbf{F}$ evaluated at the equilibrium the fixed point $\mathbf{z}^{\star}_{i}$ by $\mathbf{J}_{i}$. Then, $\mathbf{z}^{\star}_{i}$ is stable if all eigenvalues $\{ \nu^{1}_{i} \}$ and  $\{ \nu^{2}_{i} \}$ of $\mathbf{J}_{i}$ satisfies $\textrm{Re} \left(\nu^{1,2}_{i} \right) < 0$, and unstable if at least one eigenvalue satisfies $\textrm{Re} \left( \nu^{1,2}_{i} \right) > 0$. 
As a subset of the unstable systems, the cases where $\textrm{Re} \left(\nu^{2}_i \right) > 0$ and $\textrm{Re} \left(\nu^{1}_i \right) < 0$ are usually called semistables \cite{kuznetsov2013elements}.

The explicit expression for Jacobian matrix related to the vector field $\mathbf{F}$ in terms of the function $f$ is
\begin{equation}
    \mathbf{J}_{i} =
    \begin{pmatrix}
    0 & 1 \\
    \tilde{g}'(r) & 0
    \end{pmatrix}_{(r^{\star}_{i},0)} \,\, ,
    \label{jac}
\end{equation}
where $\tilde{g}(r)$ is defined in Eq.~\eqref{def-g-h}. The eigenvalues of $\mathbf{J}_{i}$ are promptly calculated
\begin{equation}
\nu^{1,2}_{i} =\pm\sqrt{\tilde{g}'(r_{i}^{\star})} \,\, . 
    \label{local_lyapunov}
\end{equation}
Result~\eqref{local_lyapunov} implies that the null-geodesic dynamics generates three possible scenarios: a semistable (and hence unstable) neutral-saddle regime when $\tilde{g}'(r_{i}^{\star}) > 0$, a marginally stable center regime when $\tilde{g}'(r_{i}^{\star}) < 0$, and a degenerated fixed point when $\tilde{g}'(r_{i}^{\star}) = 0$.

It should be noticed that, when $\tilde{g}'(r^{\star}_i) < 0$ (degenerated fixed-point case), the linear stability analysis is inconclusive, since the necessary conditions to the Hartman-Grobman theorem fail. However, with the the direct Lyapunov method, it can be checked whether the nonlinearities of the system do not destroy possible centers.
To this purpose, the direct Lyapunov method will be employed here. This method is based on the existence of a positive-definite function $V(\mathbf{z})$ which is called a Lyapunov's function. 
If the fixed point $\mathbf{z}^{\star}_{i}$ of the system~\eqref{dynamical_system_ger} is a minimum of the Lyapunov function, that is, the conditions $V(\mathbf{z}^{\star}_{i}) = 0$ and $V(\mathbf{z}) \neq 0$ for $\mathbf{z} \neq \mathbf{z}^{\star}_{i}$ are satisfied and its derivative with respect to $\eta$ is $\left. \textup{d} V / \textup{d} \eta \right|_{\mathbf{z}^{\star}_{i}} \leq 0$, then the fixed point $\mathbf{z}^{\star}_{i}$ is said to be stable. 
Asymptotic stability occurs when the derivative is strictly negative, $\left. \textup{d} V / \textup{d} \eta \right|_{\mathbf{z}^{\star}_{i}} < 0$. 

Let us consider the following quantity:
\begin{equation}
    V(r,w)=\frac{r^2 B(r)-r^2 w^2 A(r)}{A(r) B(r)} = g(r) - w^{2} h(r) \,\, ,
\end{equation}
with the functions $g(r)$ and $h(r)$ defined as
\begin{equation}
g(r) \equiv \frac{r^2}{A(r)} \,\, ,
\,\,\, 
h(r) \equiv \frac{r^2}{B(r)} \,\, .
\label{fun_h}
\end{equation}
It is straightforward to check that the orthogonality condition is satisfied for this function \cite{abolghasem2013stability}:
\begin{equation}
    \frac{\textup{d} V}{\textup{d} \eta} = \mathbf{F}(\mathbf{z}) \cdot \mathbf{\nabla} V = 0  \,\, .
\label{dvdeta}
\end{equation}
Therefore, the proposed function $V$ is a suitable candidate for a Lyapunov function.
The last condition to show is that the fixed point $r^{\star}_{i}$ is a local minimum of $V(r,w)$, thus, evaluating the Hessian matrix we obtain
\begin{equation}
    \mathbf{H}_{i} = 
    \begin{pmatrix}
     g''(r)& 0 \\
     0 & -2h(r) 
    \end{pmatrix}_{(r^{\star}_{i},0)} \,\, .
\label{def-H}
\end{equation}
From result~\eqref{def-H},
there is a local maximum if $g''(r^{\star}_i) < 0$ and $h(r^{\star}_i) > 0$, a local saddle if $g''(r^{\star}_i)> 0$ and $h(r^{\star}_i) > 0$ and a local minimum if $g''(r^{\star}_i) > 0$ and $h(r^{\star}_i) < 0$. 
Also, inspecting Eq.~\eqref{fun_h}, we notice that $h(r)$ is a positive-definite function, and therefore there is no center in the null-geodesic dynamics.
When either $g''(r^{\star}_i) = 0$ or $h(r^{\star}_i) = 0$ the stability analysis presented here is inconclusive, and a complementary approach is needed. 

Writing the functions $\tilde{g}(r)$ and $\tilde{h}(r)$ in terms of $g(r)$, $h(r)$ and their derivatives,
\begin{equation}
\tilde{g}(r) = \frac{g'(r)}{2h(r)} \,\, , \,\,\, 
\tilde{h}(r) = - \frac{h'(r)}{h(r)} \,\, .
\label{def-g-h}
\end{equation}
Rewriting the Hessian matrix~\eqref{def-H} in terms of $\tilde{g}(r)$ we get 
\begin{equation}
    \mathbf{H}_{i} = 
    \begin{pmatrix}
     2\tilde{g}'(r)h(r)& 0 \\
     0 & -2h(r) 
    \end{pmatrix}_{(r^{\star}_{i},0)} \,\, .
\label{def-H_tilde}
\end{equation}
\\
Hence the stability of the system according to the Lyapunov function is characterized by the sign of $\tilde{g}'(r_{i}^{\star})$. If $\tilde{g}'(r_{i}^{\star}) < 0$, the fixed point is a nonhyperbolic semistable saddle. If $\tilde{g}'(r_{i}^{\star}) > 0$, the fixed point is a semistable neutral saddle. 
That is, the development presented so far indicates that the indirect and direct Lyapunov methods are consistent.
However, if $\tilde{g}'(r_{i}^{\star}) = 0$, the fixed point is an inflection point in the Lyapunov function and the definition of stability cannot be determined with this method.

Another important characterization of stability is the Jacobi criterion.%
\footnote{The connection between Jacobi and linear stability was previously studied in detail by several authors \cite{sabau2005some,boehmer2012jacobi,abolghasem2013stability}.}
To introduce the concept of Jacobi stability, we will outline the method in the following, restricting the formalism to the CFM case. Consider the second-order differential equations in the standard form \cite{sabau2005some},
\begin{equation}
    \frac{\textup{d}^{2} x}{\textup{d}\eta^{2}} +2G(x,\dot{x}) = 0  \,\, ,
    \label{second_order_system}
\end{equation}
with a fixed point at  $x=0$.
Let us denote by $\xi$ the first-order variation of $x$.
Defining the Kosambi-Cartan-Chern-covariant derivative of the field $\xi^{i}$ \cite{sabau2005some,boehmer2012jacobi},
\begin{equation}
    \frac{D\xi}{\textup{d}\eta} = \frac{\textup{d}\xi}{\textup{d}\eta}+\frac{\partial G}{\partial \dot{x}}\xi \,\, ,
    \label{kcc_derivative}
\end{equation}
and varying the trajectories of~\eqref{second_order_system} into nearby ones, the Jacobi equations \cite{sabau2005some} are obtained:
\begin{equation}
    \frac{D^{2}\xi}{\textup{d}\eta^{2}} = P\xi \,\, .
    \label{jacobi_equation}
\end{equation}
The so-called deviation curvature tensor \cite{sabau2005some} (in the present case, a scalar) is defined as 
\begin{equation}
    P = -2\frac{\partial G}{\partial x}-2G \tilde{G} + \dot{x}\frac{\partial N}{\partial x} + (N)^{2} \,\, ,
    \label{deviation_tensor}
\end{equation}
where $\dot{x} \equiv \textup{d}x / \textup{d} \eta$ and the nonlinear and Berwald terms ($N$ and $\tilde{G}$ respectively) \cite{boehmer2012jacobi} are given by 
\begin{equation}
    N = \frac{\partial G}{\partial \dot{x}} \,\, , \qquad \tilde{G} = \frac{\partial N}{\partial \dot{x}} \,\, .
\end{equation}
With the relevant quantities introduced, Jacobi's criterion of stability \cite{sabau2005some,boehmer2012jacobi} can be stated. The trajectories of~\eqref{second_order_system} are Jacobi stable if $P(0)$ are strictly negative, and Jacobi unstable otherwise.

Taking the null-geodesic dynamics written in the form~\eqref{geodesic_equation}, the Berwald conection is written as
\begin{equation}
\tilde{G}(r,w) = -\frac{\tilde{f}(r,w)}{2} \,\, ,
\end{equation}
and the following quantities can be computed:
\begin{equation}
N = \frac{\partial G(r,w)}{\partial w} \,\, , \qquad \tilde{G} = \frac{\partial N}{\partial w} = \frac{\partial^{2} G(r,w)}{\partial w^{2}} \,\, .
\label{N_tildeG}
\end{equation}
From the expressions for $P$, $N$ and $\tilde{G}$ in Eqs.~\eqref{deviation_tensor} and \eqref{N_tildeG} \cite{sabau2005some}, the deviation curvature scalar is calculated:
\begin{equation}
P = -2\frac{\partial G (r,w)}{\partial r} - 2G\frac{\partial^{2} G}{\partial w^{2}} + w \frac{\partial G}{\partial r \partial w} +\left( \frac{\partial G}{\partial w}\right)^{2} \,\, .
\end{equation}
Noticing that $w = 0$ and $G (r^{\star}_{i}, 0) = 0$ for the photon sphere, it follows that
\begin{equation}
P = -2\frac{\partial G}{\partial r} + \left( \frac{\partial G}{\partial w}\right)^{2} \,\, .
\end{equation}
Comparing with the Jacobian in Eq.~\eqref{jac} we get
\begin{equation}
P = -\det(\mathbf{J}) + \frac{1}{4}\, [\textrm{tr}(\mathbf{J})]^{2}
= \tilde{g}'(r^{\star}_{i}) \,\, .
\label{P1} 
\end{equation}
Result~\eqref{P1} indicates that the system is Jacobi unstable for $\tilde{g}'(r_{i}^{\star}) > 0$.

We conclude that the stability analysis using Jacobi criterion coincides with the results obtained from Lyapunov's criteria (direct and indirect methods). This is a significant result, since linear (Lyapunov) and Jacobi stability criteria are not in agreement considering many examples of dynamic systems \cite{abolghasem2013stability,sabau2005some,boehmer2012jacobi}. Indeed, there is a special interest in the scenarios where the coexistence of both stabilities can be checked, which indicates a robust behavior of the system \cite{sabau2005some}.

\section{\label{sec:bifurcations} Bifurcations}

In a broad sense, a bifurcation in a dynamical system occurs when a small variation in the parameters characterizing the system generates a qualitative change in the behavior of the system. We will consider this issue in the dynamics of null geodesics, and in the presented discussion the possibility of semistable photon spheres in the CFM spacetimes will be shown.
The bifurcation associated to the variation in the number of photon spheres can be found considering the condition $r^{\star}_{1} = r^{\star}_{2}$, that is,
\begin{equation}
3M = \frac{M}{2} \left( 4 - C^{\textrm{b}} \right) \Longrightarrow C^{\textrm{b}} = -2 \,\, .
\end{equation}
Hence, if the deformation parameter is given by $C=C^{\textrm{b}} \equiv -2$, a photon sphere with degenerated null eigenvalues is produced. 

\begin{figure}[ht]
    \centering
    \includegraphics[scale=1]{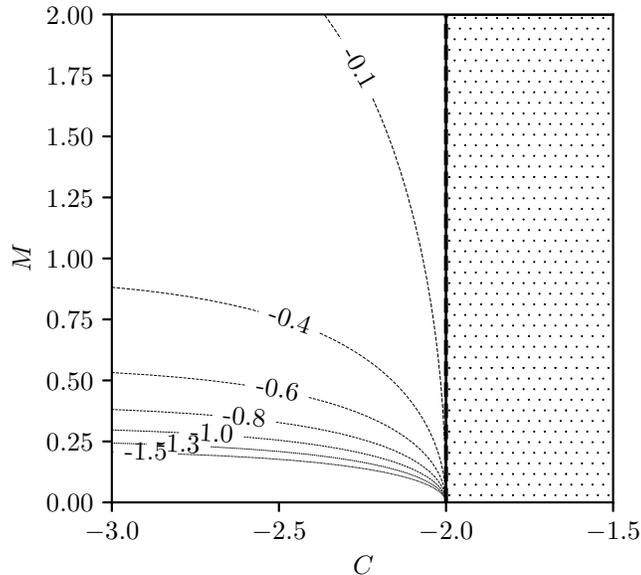}
    \caption{Hyperbolic regions for the photon sphere located at $r=r^{\star}_{1} \equiv M(4-C)/2$. The photon sphere is nonhyperbolic in the dot-filled region and hyperbolic in the contour region (showing the level curves for the magnitude of the eigenvalues $\nu^{1,2}_{1}$).}
    \label{fig:estabCE2}
\end{figure}

\begin{figure}[ht]
    \centering
    \includegraphics[width=8.6cm]{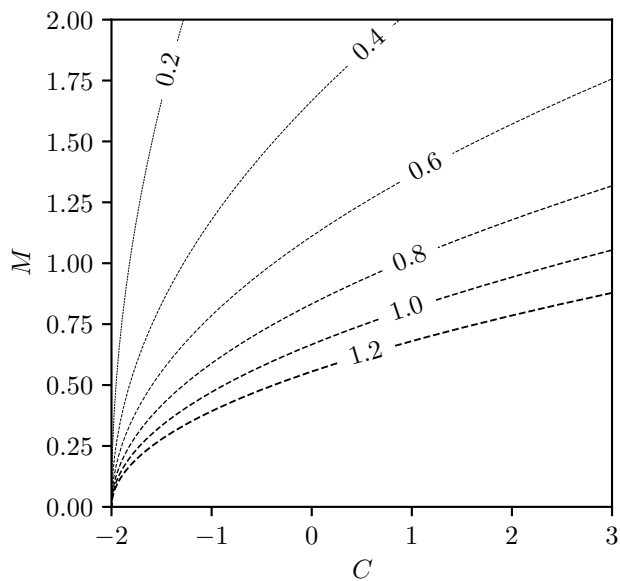}
    \caption{Instability region in the parameter space for the photon sphere located at $r=r^{\star}_{2} \equiv 3M$, with the level curves for the magnitude of the eigenvalues $\nu^{1,2}_{2}$.}
    \label{fig:estabCE1}
\end{figure}

Evaluating the eigenvalues $\nu^{1}_{1}$ and $\nu^{2}_{1}$ of the Jacobian, considering the photon sphere localized at $r = r^{\star}_{1}$,
\begin{equation}
\nu^{1,2}_{1} = \pm \frac{2 (C+2)}{\sqrt{(C-4) (C-1) C (C+2)} M}  \,\, . 
\label{first-phot}
\end{equation}
Result~\eqref{first-phot} indicates that the mass $M$ of the compact body does not influence the stability of this photon sphere. Still, variations in $M$ can increase or decrease the magnitude of the eigenvalues. We illustrate this point in figure~\ref{fig:estabCE2}, where the contour plot associated to Eq.~\eqref{first-phot} is presented. 
In this figure, the nonhyperbolic region is marked with a dotted pattern. The photon sphere becomes degenerated in the regions marked with contours and in the dashed line $\nu^{1,2}_{1} = 0$.

By checking the critical value $D^{c}_{1}$ for the impact parameter $D$ in Eq.~\eqref{crit-par-photon-1}, we verify that $D^{c}_{1} \notin \mathbb{R}$ in the range $0 < C < 4$, meaning that the the photon sphere located at $r=r^{\star}_{1}$ is not the asymptotic limit of a null geodesic.
In short, for $C > 0$ it follows that $r^{\star}_{1}$ is not associated to a photon sphere and for $C < -2$ the photon sphere at $r=r^{\star}_{1}$ is unstable.

Considering the photon sphere located at $r = r^{\star}_{2}$, the associated eigenvalues $\nu^{1}_{2}$ and $\nu^{2}_{2}$ are given by
\begin{equation}
\nu^{1,2}_{2} = \pm \frac{\sqrt{(C+2)}}{3M} \,\, .
\label{photo-3M}
\end{equation}
Result~\eqref{photo-3M} shows that the photon sphere at $r=r^{\star}_{2}$ is hyperbolic for $C > -2$. 
This point is illustrated in figure~\ref{fig:estabCE1}.
When $C = C^{\textrm{b}} \equiv -2$ we get $\nu^{1,2}_{1} = 0$ in the critical photon sphere located at $r=r^{\star}_{1}$, as can be seen from Eq.~\eqref{fixed-points}. This result indicates that a Bogdanov-Takens bifurcation is present when $C = -2$. This bifurcation is associated with the coalescence of the two photon spheres, destroying the heteroclinic orbit, as it will be commented in the next section.
When $C < -2$, we have seen in section~\ref{sec:photon-spheres} that $r^{\star}_{2}$ is not associated to a photon sphere.

\begin{figure}[ht]
    \centering
    \includegraphics[scale=1]{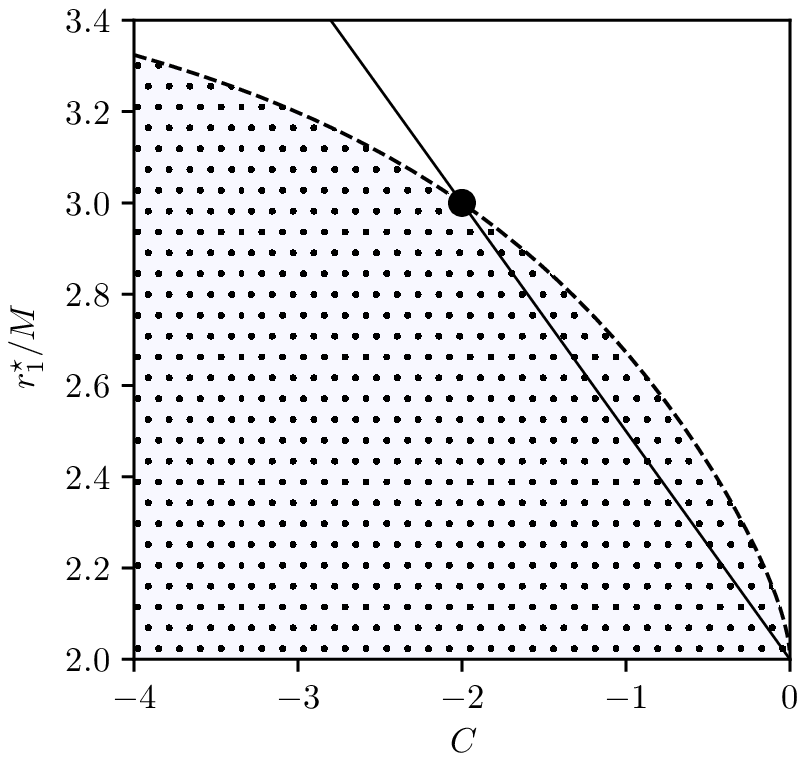}
    \includegraphics[scale=1]{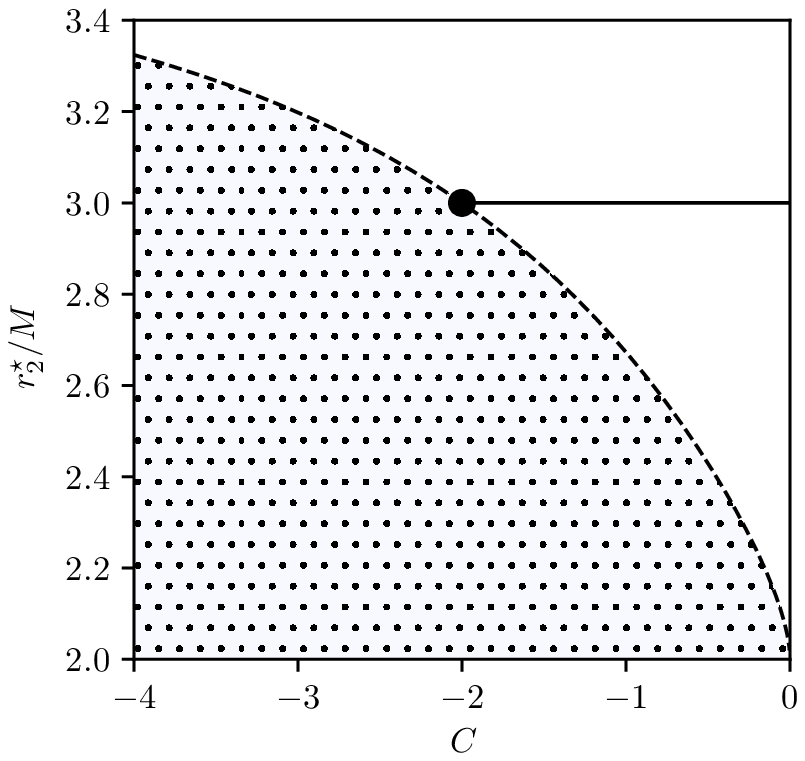}
    \caption{Photon-sphere manifold for the photon spheres located at $r=r_{1}^{\star}$ and $r=r_{2}^{\star}$ (left and right panels). The dot-filled regions indicate hyperbolicity and the clear regions indicate nonhyperbolicity (both related to the sign of $\tilde{g}'(r^\star_{i})$), while in the boundary between the regions (dashed lines) the Jacobian eigenvalues are zero. Superimposed with the photon-sphere manifolds are the functions $r^{\star}_{1}/M$ (left panel) and  $r^{\star}_{2}/M$ (right panel) in term of $C$. The bifurcation point $C=C^{\textrm{b}}\equiv -2$ is marked with large dots in both panels.}
\label{fig:estabD}
\end{figure}

In order to better characterize the bifurcations in the null-geodesic system, we consider the photon-sphere manifold%
\footnote{In the literature of dynamical systems this is called ``equilibrium manifold,'' but since we are dealing with photon spheres it is suitable to use the terminology ``photon-sphere manifold.''}
in the $(C, r^{\star}_{i})$ space,
defined by \cite{kuznetsov2013elements}
\begin{gather}
\tilde{f}(r^{\star}_{i},0) = 0 
\label{eq:bif_mani_1}
\end{gather}
for bifurcations related to the number of photon spheres, and 
\begin{gather}
    \tilde{f}'(r^{\star}_{i},0) = 0 
\label{eq:bif_mani_2} 
\end{gather}
for bifurcations related to the change of hyperbolicity on the fixed points. 
The manifold defined by~\eqref{eq:bif_mani_1} is illustrated in figure~\ref{fig:estabD}. 
In this figure, if $C > C^{\textrm{b}}\equiv -2$ , the photon sphere at $r=r^{\star}_{1}$ is nonhyperbilic and the photon sphere at $r=r^{\star}_{2}$ is hyperbolic.
When $C < C^{\textrm{b}}$, there is only one unstable photon sphere at $r=r^{\star}_{1}$. Heuristically, when the two photon spheres ``collide'' in the limit $C \rightarrow -2^{+}$, one of then is annihilated.

Since it is necessary to vary two parameters of the system in order to qualitatively change  the system, there is a codimension-two bifurcation associated to the null-geodesic dynamics in CFM spacetimes. Specifically, this bifurcation produces a degenerated photon sphere located at $r = 3M = r^{\star}_{1} = r^{\star}_{2}$, when
\begin{equation}
    C = C^{\textrm{b}} \equiv -2 \,\, .
\end{equation}
Evaluating the Jacobian matrix in the bifurcation point, one obtains
\begin{equation}
\mathbf{J} = 
\begin{pmatrix}
0 & 1 \\
0 & 0
\end{pmatrix} \,\, ,
\label{Jordan-block}
\end{equation}
which is the zero Jordan block of order $2$ with a doubly degenerate null eigenvalue. 
It implies the existence of a Bogdanov-Takens bifurcation \cite{kuznetsov2013elements,kuznetsov2005practical}. 

Linear stability analysis is not sufficient for the proper characterization of this case, since both eigenvalues are on the central manifold of the system. In this scenario, to study the stability of the bifurcation point, we will evaluate the coefficients of the normal form of the Bogdanov-Takens bifurcation \cite{kuznetsov2005practical}.

Near a fixed point, the restriction of the field $\mathbf{F} (\mathbf{z})$ in Eq.~\eqref{dynamical_system_ger} to any center manifold in the critical parameter values can be expressed by the polynomial approximation $\mathbf{z} = \mathbf{H}(\zeta_0,\zeta_1)$, specifically,
\begin{equation}
H(\zeta_0,\zeta_1) = \zeta_0q_0 +\zeta_1q_1 + \sum_{2 \leq j+k \leq 4} \frac{1}{j!k!}h_{jk}\zeta_{0}^{j}\zeta^{k}_{1} + \mathcal{O}(||(\zeta_0,\zeta_1)||^5) \,\, ,
\label{approx}
\end{equation} 
where $q_0$ and $q_1$ are the eigenvectors of $\mathbf{J}_i$.  From~\eqref{approx}, the invariant central manifold conditions furnish the associated homological equation,
\begin{equation}
\frac{\partial H}{\partial \zeta_0}\frac{\textup{d}\zeta_0}{\textup{d}\eta} + \frac{\partial H}{\partial \zeta_1}\frac{\textup{d}\zeta_1}{\textup{d}\eta} = F(H(\zeta_0,\zeta_1)) \,\, .
\label{homo_eq}
\end{equation}
With a coordinate change \cite{kuznetsov2005practical}, Eqs.~\eqref{homo_eq} and~\eqref{approx} yield the normal form
\begin{eqnarray}
\frac{\textup{d}\zeta_0}{\textup{d}\eta} &=& \zeta_1 \,\, , \\
\frac{\textup{d}\zeta_1}{\textup{d}\eta} &=& \sum_{k \geq 2} (a_{k} \zeta_0^{k} + b_{k}\zeta_0^{k-1}\zeta_1) \,\, .
\end{eqnarray} 
The coefficients $a_{k}$ and $b_{k}$ depend on $\mathbf{F}(\mathbf{z})$.%
\footnote{For more details about the computation of normal forms of Bogdanov-Takens bifurcation with codimension 2 and 3, see for example \cite{kuznetsov2005practical}.}
Direct evaluation of the coefficients furnishes $b_{2} = 0$ \cite{kuznetsov2005practical}, indicating that the system exhibits a degenerated Bogdanov-Takens bifurcation, whose normal form is
\begin{eqnarray}
  \frac{\textup{d}\zeta_{0}}{\textup{d} \eta} & = & \zeta_{1} \,\, ,  
\label{normal-form-1}
\\
   \frac{\textup{d}\zeta_{1}}{\textup{d} \eta} & = & a_{2} \zeta_{0}^{2} + b_{4}\zeta_{0}^{3}\zeta_{1} + \mathcal{O}(||(\zeta_{0},\zeta_{1})||^{5}) \,\, ,
\label{normal-form-2}
\end{eqnarray}
where $a_{2}$ and $b_{4}$ are, respectively, the second-order and fourth-order coefficients of the normal form. Evaluating these coefficients \cite{kuznetsov2005practical}, we verify that $a_{2}=1/3$ and $b_{4} = 0$. Hence, close to the bifurcation, the phase portrait of the dynamical system~\eqref{vec-CFM-1-eff}--\eqref{vec-CFM-2-eff} is locally topologically equivalent to the system \cite{kuznetsov2005practical}:
\begin{eqnarray}
    \frac{\textup{d}\zeta_{0}}{\textup{d}\eta} & = & \zeta_{1} \,\, ,  
     \label{normal_form_system-1}\\
    \frac{\textup{d}\zeta_{1}}{\textup{d}\eta} & = & \frac{1}{3}\zeta_{0}^{2} + \mathcal{O}(||(\zeta_{0},\zeta_{1})||^{5}) \,\, .      
     \label{normal_form_system-2}
\end{eqnarray}

\begin{figure}[ht]
    \centering
    \includegraphics[scale=1]{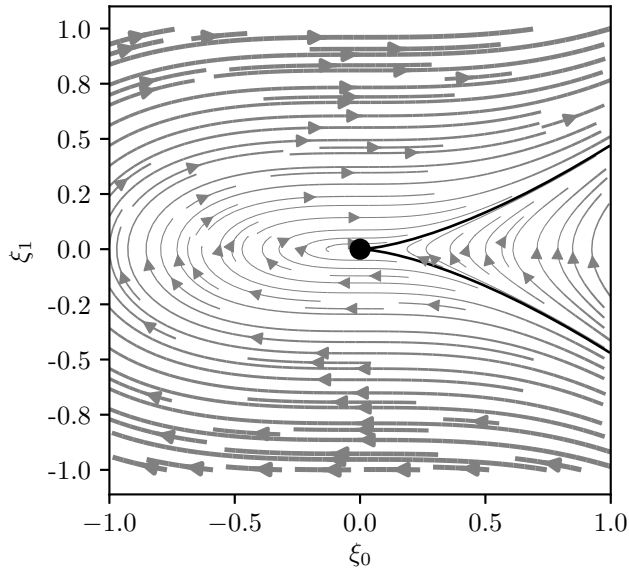}
    \caption{Phase portrait of the normal form~\eqref{normal_form_system-1}--\eqref{normal_form_system-2} of the Bogdanov-Takens bifurcation, with a cusp at the origin, associated to a bifurcation of the null-geodesic dynamics in CFM spacetimes.}
    \label{fig:cusp}
\end{figure}

Considering the normal form~\eqref{normal_form_system-1}--\eqref{normal_form_system-2}, the associated Jacobian is
\begin{equation}
    \mathbf{J} = \begin{pmatrix}
    0 & 1 \\
    1/3 & 0
    \end{pmatrix} \,\, .
\label{new-J}
\end{equation}
The eigenvalues of $\mathbf{J}$ in Eq.~\eqref{new-J} are given by $\nu^{1,2} = \pm\sqrt{1/3}$ and therefore the phase portrait displays an unstable cusp point in the critical photon sphere ($r = r^{\star}_{1} \equiv 3M$ and $C = C^{\textrm{b}} \equiv -2$), as illustrated in figure~\ref{fig:cusp}.

\section{\label{sub:heteroclinic} Heteroclinic orbits}

Of considerable importance in a dynamical system is the possible existence of heteroclinic orbits. A heteroclinic orbit joins two different fixed points. These structures are present in CFM spacetimes, specifically considering wormhole backgrounds ($C<0$) and extreme black holes ($C=0$), as it will be discussed in this section.

\begin{figure}[ht]
    \centering
    \includegraphics[width=8.6cm]{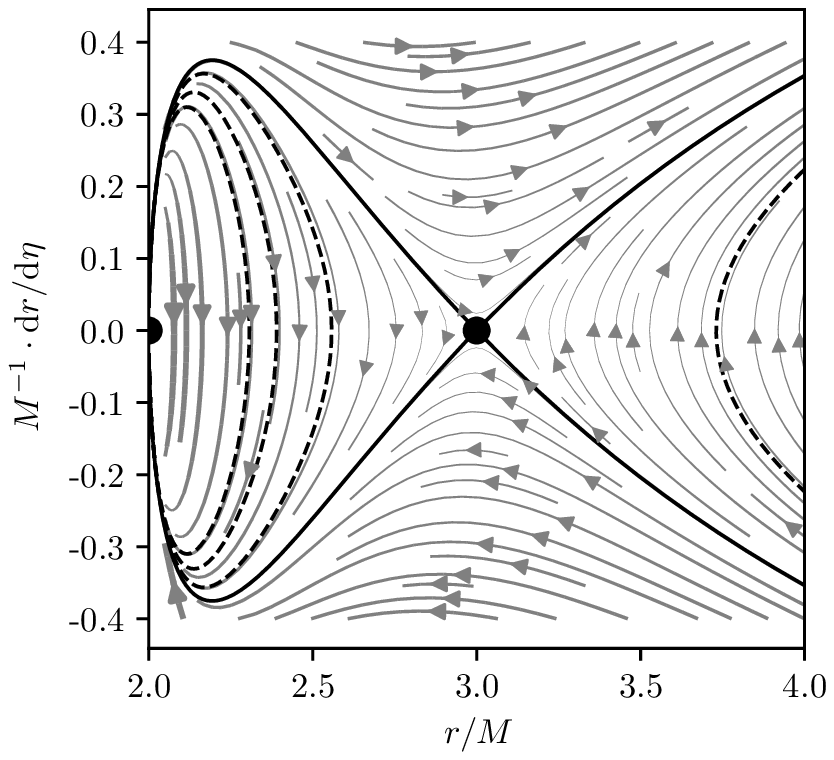}
    \includegraphics[width=8.6cm]{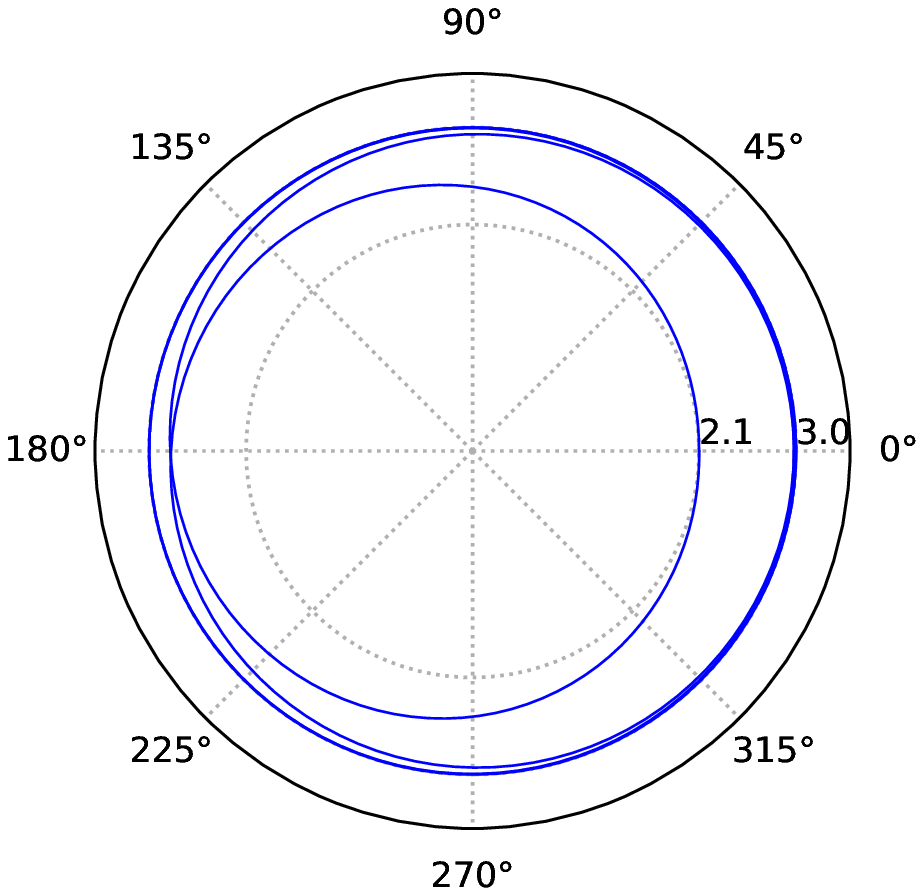}
     \caption{The emergence of a bounded orbit (left dashed lines) and heteroclinic orbit (left continuous loop) in the phase portrait and in the orbital plane, in the left and right panels, respectively. The parameters for the left graph are $C = 0$, $M = 1$ and for the right graph are $C = -0.2$, with initial conditions $(r(0),w(0)) = (2.11, 0.00)$.}
    \label{fig:heteroclinic}
\end{figure}

For values of the deformation parameter in the range $-2 < C \leq 0$, heteroclinic orbits are present. A geodesic  originated at the photon sphere with radius $r_{2}^{\star}$ goes until the throat and then returns back. This path forms a heteroclinic loop, which is illustrated in figure~\ref{fig:heteroclinic}. 
As the parameter $C$ approaches $-2$ from the right, this family of bounded orbit contracts until it disappears when $C = C^{\textrm{b}} \equiv -2$, giving rise to a Bogdanov-Takens bifurcation.
The maximum amplitude of the orbits in this set is $r = M$ and it occurs when $C \to 0^{-}$. It is possible to access the region of bounded orbits by conveniently adjusting the initial conditions of the geodesics.

The fixed point at the wormhole throat is a nonlinear saddle, while the other fixed point is a neutral saddle. Those, in turn, generate heteroclinic loops, which are the closed orbits connecting the saddle fixed points. After the Bogdanov-Takens bifurcation, this region disappears and the system admits just scattering states.

The existence of heteroclinic orbits in wormhole configurations has important consequences. Following the Peixoto's theorem \cite{peixoto1959structural}, heteroclinic orbits are associated to structural instabilities. That is, the variation of any system's parameter can destroy these special orbits and then change the topology of the phase portrait. 
This phenomenon is related to the emergence of unbounded orbits (scattering states), and consequently, the inaccessibility of the stable photon sphere in the orbital plane for critical geodesics. These unbounded orbits correspond to unstable orbits, as can be seen from result~\eqref{first-phot}.

\section{\label{sec:conclusions} Final remarks}

An extensive dynamical analysis of the null geodesics in the spacetimes proposed by Casadio, Fabbri, and Mazzacurati was performed, considering both black hole and wormhole solutions. Stability was studied within the Lyapunov and Jacobi framework.
Bifurcations and the presence of heteroclinic orbits were investigated. It was observed a direct relation between the null geodesics and the structure of the spacetimes, which  gives us a signature of the inner geometry of the black holes, and so could be used as future observational tests. The connections between the null-geodesic dynamics and the associated CFM spacetimes are summarized in table~\ref{tab:Spacetimegeodesics}.

\begin{table}[h]

    \caption{Association between the characteristics of the CFM spacetimes and the null-geodesic dynamics.}
  
\begin{tabularx}{\textwidth}{YYYY}
\hline \hline
\begin{minipage}[c]{0.15\textwidth}
\textbf{Deformation parameter} 
\end{minipage}
& 
\textbf{Compact structure} 
& 
\textbf{Null-geodesic dynamics}  
& 
\begin{minipage}[c]{0.20\textwidth}
\mbox{} \\ \textbf{Position of the photon spheres} \\ \mbox{}
\end{minipage}
\\
\hline
$ C < -2$ & 
Wormhole & 
\begin{minipage}[c]{0.25\textwidth}
\mbox{} \\ One semistable and nonhyperbolic photon sphere at $r=r^{\star}_{1}$. \\ \mbox{} 
\end{minipage}
& 
\begin{minipage}[c]{0.2\textwidth}
$r^{\star}_{1} > 3M$ \\ 
\mbox{}
\end{minipage}
\\
\hline
$ C = -2$ &
Wormhole &
\begin{minipage}[c]{0.25\textwidth}  
\mbox{} \\ Bogdanov-Takens bifurcation. One degenerate, semistable and nonhyperbolic photon sphere at $r=r^{\star}_{1}=r^{\star}_{2}$. \\ \mbox{}  
\end{minipage}
& 
$r^{\star}_{1}=r^{\star}_{2}=3M$ \\
\hline
$ -2 < C < 0$ &  
Wormhole &  
\begin{minipage}[c]{0.25\textwidth}
\mbox{} \\ Bounded heteroclinic orbits. One semistable and nonhyperbolic photon sphere at $r = r^{\star}_{1}$. Two semistable and hyperbolic photon spheres at $r = r^{\star}_{2}$ (one on each side of the wormhole throat). \\ \mbox{}
\end{minipage}
&
\begin{minipage}[c]{0.2\textwidth}
$2M < r^{\star}_{1} < 3M$ \\ 
\mbox{} \\
$r^{\star}_{2} = 3M$ 
\end{minipage}
\\
\hline
$C = 0$ &
Extreme black hole &
\begin{minipage}[c]{0.25\textwidth}
\mbox{} \\ Bounded orbits near the event horizon, one semistable and hyperbolic photon sphere at $r=r^{\star}_{2}$. 
\\ \mbox{}
\end{minipage}
&
$r^{\star}_{2} = 3M$ \\
\hline
$C>0$ &
Nonextreme black hole &
\begin{minipage}[c]{0.25\textwidth}
\mbox{} \\ One semistable and hyperbolic photon sphere at $r=r^{\star}_{2}$. \\ \mbox{}
\end{minipage}
&
$r^{\star}_{2} = 3M$ \\
\hline
\hline
    \end{tabularx}
    \label{tab:Spacetimegeodesics}
\end{table}

The theory concerning static spherically symmetric black holes is well developed. For instance, it is known that in several cases their photon spheres are unstable under small perturbations \cite{chandrasekhar1985mathematical,Khoo:2016xqv}. In the present work, this characteristic was observed for both photon surfaces of the system. The photon sphere located at $r= r^{\star}_{2} \equiv 3M$ displays hyperbolicity for all black-hole regimes, losing this feature just when $C = -2$. Furthermore, the photon surface at $r=r_1^{\star}$ is always nonhyperbolic.

We explored these issues, showing that both Lyapunov and Jacobi stability criteria are concordant in the null-geodesic dynamics considering the CFM solutions. A novel feature of the CFM spacetimes is the loss of hyperbilicity of the photon surfaces as the parameter $C$ is varied. This leads to a richer dynamics for this family of solutions, when comparing with the usual Schwarzschild case.

It was also shown that there is a bifurcation taking place when the parameter $C$ is varied. This bifurcation was classified as a degenerated Bogdanov-Takens. Therefore, this bifurcation can be thought as a combination of a fold and transcritical bifurcations, which displays an unstable cuspid point at the critical photon sphere. These topological changes can be observed in the orbital plane as the deformation parameter is varied. The bifurcation gives rise to a region of bounded states (orbits which are limited to a finite portion of the orbital plane) where a semistable photon sphere exists when the deformation parameter is in the range $-2 < C < 0$. 
It should be remarked that a fold bifurcation was observed in Friedmann-Lema{\^ i}tre-Robertson-Walker scenarios \cite{kohli2018einstein}. That is, we are presenting similarities between the local physics associated to compact objects and global physics related to a cosmological solution. We speculate that the results presented here may reveal unseen connection between apparently nonrelated gravitational systems.

Heteroclinic orbits were found in the CFM null-geodesics dynamics. They are the separatrix between the region of bound and scattering states of this system. When the system goes through the Bogdanov-Takens bifurcation this separatrix shrinks until it disappears and only scattering states are allowed in the orbital plane. The wormhole case showed the presence of a heteroclinic loop that separates the regions of bounded and scattering states, enclosing a semistable photon sphere and connecting two unstable photon spheres. 
Finally, our results suggest that CFM black-hole spacetimes show structural stability, that is, the topology of their phase portrait does not change under parameter variation. On the other hand, the wormhole spacetimes do not appear to be structurally stable, as a consequence of the existence of heteroclinic orbits and Peixoto's theorem \cite{peixoto1959structural}. Further investigations on this topic are being conducted.

\newpage

\begin{acknowledgments}

We thank Alberto Saa, Iber\^{e} Luiz Caldas and Masayuki Oka Hase for the helpful comments.
W.S.K. acknowledges the support of the Coordination for the Improvement of Higher Education Personnel (CAPES), Brazil, Finance Code 001.
C.M. acknowledges the support of National Council for Scientific and Technological Development (CNPq), Brazil, Grant No. 420878/2016-5.
\end{acknowledgments}

\end{document}